\documentclass[twocolumn, showpacs, preprintnumbers, amsmath, amssymb, superscriptaddress, pra]{revtex4}
\usepackage{graphicx}
\usepackage{dcolumn}% Align table columns on decimal point
\usepackage{bm}% bold math
\usepackage{epsf}
\usepackage{amssymb}
\DeclareGraphicsExtensions{.eps}
\begin{document}

\author{D.S. Simon}
\affiliation{Dept. of Electrical and Computer Engineering, Boston
University, 8 Saint Mary's St., Boston, MA 02215}
\affiliation{Photonics Center, Boston
University, 8 Saint Mary's St., Boston, MA 02215}
\author{A.V. Sergienko}
\affiliation{Dept. of Electrical and Computer Engineering, Boston
University, 8 Saint Mary's St., Boston, MA 02215}
\affiliation{Photonics Center, Boston
University, 8 Saint Mary's St., Boston, MA 02215}
\affiliation{Dept. of Physics, Boston University, 590 Commonwealth
Ave., Boston, MA 02215}

\begin{abstract} We discuss a correlated two-photon imaging apparatus that is capable of producing
images that are free of the effects of odd-order
aberration introduced by the optical system. We show that both quantum-entangled and classically
correlated light sources are capable of producing the desired
spatial-aberration cancellation.
\end{abstract}

\title{Odd-Order Aberration-Cancellation in Correlated-Photon Imaging}

\pacs{42.30.Va,42.15.Fr,42.50.-p}

\maketitle

\section{Introduction}

Two-photon imaging, also known as ghost imaging, involves the
use of coincidence measurements to form
images via photons that never interacted with the object being
viewed. It has been a topic of great interest since its discovery
using entangled photon pairs \cite{pittman}. Initially, it was
believed that the entanglement was a necessary ingredient for
the effect, but it has since been
found that most aspects of ghost imaging can be simulated using
spatially-correlated classical light \cite{bennink1, bennink2},
including thermal and speckle sources \cite{gatti, cai, valencia,
scarcelli, ferri, zhang}.

Separately, it has been demonstrated that the entangled photon
pairs produced in spontaneous parametric downconversion (SPDC) may
also be used to cancel some of the effects of frequency dispersion
\cite{steinberg, franson, minaeva1} or spatial dispersion
(aberration) \cite{bonato1, bonato2, simon1} in
interferometer experiments.

The goal here is to show that two-photon imaging can be done with spatially correlated
pairs of light beams in such a way that odd-order aberration effects introduced by an imaging
system may be cancelled. This may be
done with either classically correlated beams or quantum entangled photon pairs. The key idea
is to partially collapse the two arms of the ghost-imaging setup so that {\it both} of the output
beams pass through the {\it same} optical
system in an anticorrelated manner, but with only one of the beams passing through the object.
Here we will illustrate the method using the simplest
possible imaging system, a single lens.

The outline of the paper is as follows. In section II, we review
the basic idea of correlated two-photon imaging. In section III we
discuss aberration-cancelled two-photon imaging with an entangled light
source. In section IV we briefly discuss how to do the same with a classical source,
followed by conclusions in section V.

\section{Two-Photon Ghost imaging}

Two-photon correlated imaging, or ghost imaging \cite{pittman}, is done with an apparatus like the
one depicted schematically in figure 1. In the original version,
the correlated photon source is a $\chi^{(2)}$ nonlinear crystal
pumped by a laser, leading to spontaneous parametric
downconversion. This is the source we will use in section III.
Entangled photon pairs with anticorrelated
transverse spatial momenta ${\mathbf q}$ and $-{\mathbf q}$ travel
along the two arms of the apparatus. The object to be viewed
introduces a modulation (either the transmittance or reflectance
of the object), given by a function $G({\mathbf x})$. This object is
placed in arm 2 (the upper branch), followed by a large bucket
detector, $D_2$. The detector's area is integrated over, so $D_2$
can not record any information on the position or momentum of the
photon that reached the object; all this detector is able to tell
us is whether the photon reached the detector unimpeded, or
whether its passage was blocked by the object. In the other branch
of the apparatus there is no object, and all of the photons reach
an array of pointlike detectors or some other form of spatially
resolving detector without hindrance. A
coincidence circuit is used to record a count every time a photon
detection occurs simultaneously (within the coincidence time
window) at both detectors. By plotting the coincidence rate as a
function of position $x_1$ in detector 1, we build up an image of
the object. This is true even though photons that actually
encountered the object in branch 2 left no record of the object's
position, and the photons in branch 1 that do carry position
information never encounter the object.

\begin{figure}
\centering
\includegraphics[totalheight=2.0in]{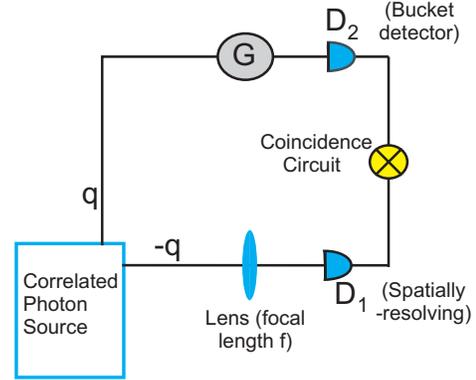}
\caption{\textit{(Color online) Schematic depiction of two-photon (ghost) imaging setup used to
view object $G$.}}
\end{figure}

\begin{figure}
\centering
\includegraphics[totalheight=2.0in]{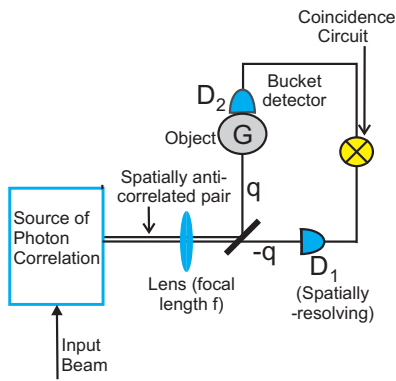}
\caption{\textit{(Color online) Schematic depiction of setup for odd-order aberration-cancelled imaging of object $G$.}}
\end{figure}

The essential ingredient is the spatial correlation of the
downconverted photon pair. The question arose as to whether the
entanglement of the photons was necessary, or if a classical
source with anticorrelated transverse momenta could mimic the
effect. It was found \cite{bennink1, bennink2} that this was
indeed possible. The correlated light source in this case consists
of a beam steering modulator (a rotating mirror, for example)
which directs a classical light beam through a range of ${\mathbf
q}$ vectors, illuminating different spots on the object. The
beamsplitter then turns the single beam of transverse momentum
${\mathbf q}$ into a pair of beams with momenta ${\mathbf q}$ and
$-{\mathbf q}$.  The results were similar to those with the
entangled source, but with half the visibility. It has since been shown
that spatial correlations present in radiation produced using
thermal and speckle sources may also lead to ghost imaging
(\cite{gatti, cai, valencia, scarcelli, ferri, zhang}).

\section{Odd-Order Aberration Cancellation in Two-Photon Imaging}

Many optical devices work by adding position-dependent phase shifts to a beam as it passes
through the device. A lens, for example, produces focusing by adding to the beam a quadratic phase shift
$e^{-{{ik}\over {2f}}{\mathbf x}^2}$, where $k$, $f$ and ${\mathbf x}$ are the wavenumber of the beam, the focal length of the lens, and the displacement of the given point in the beam from the axis of the lens.
However, imperfections in the shape of the
lens and variations in the index of refraction of the material from which it is made may lead to additional unwanted phase shifts $e^{i\phi ({\mathbf x})}$ beyond those intended. These unwanted phases are an example of optical aberration: they lead to distortions
of the outgoing wavefronts, and consequently produce distortion in the final image.
The aberration function $\phi ({\mathbf x})$ may be parameterized in a number of ways, for example by expanding in Zernike or Seidel polynomials \cite{bornwolf,buchdahl,wyant}.
Here we are uninterested in the details of how the function is represented and are concerned only with the fact
that it may be split into a sum of parts which are either even or odd under reflection about the axis:\begin{equation}\phi({\mathbf x}) =
\phi_{even}({\mathbf x})+\phi_{odd}({\mathbf x}),\end{equation} where
\begin{eqnarray} \phi_{even}(-{\mathbf x})&=&\phi_{even}({\mathbf x}),\\
\phi_{odd}(-{\mathbf x})&=&-\phi_{odd}({\mathbf x}).\end{eqnarray}
The even-order terms include astigmatism and spherical aberration, while coma
contributes to the odd-order terms.

We now wish to take the two-photon imaging setup of fig. 1 and alter
it in order to cancel as much of the aberration of the lens as possible.
Note that we should aim only to cancel the odd part of the aberration,
since cancellation of the even-order phases would also completely
cancel the effect (the quadratic phase shift) of the lens itself.

So consider the setup shown in fig. 2. This is similar to fig. 1, but with one main difference:
we have partially merged the two arms, separating the photons in each pair only {\it after} they have
passed through the lens.
The two photons both arise from the same well-localized point ${\mathbf \xi}$ in the source,
but emerge with opposite transverse momenta $\pm {\mathbf q}$. So {\it both} pass through
the same lens, but tend to pass through it on opposite sides of the axis.

\begin{figure}
\centering
\includegraphics[totalheight=2.0in]{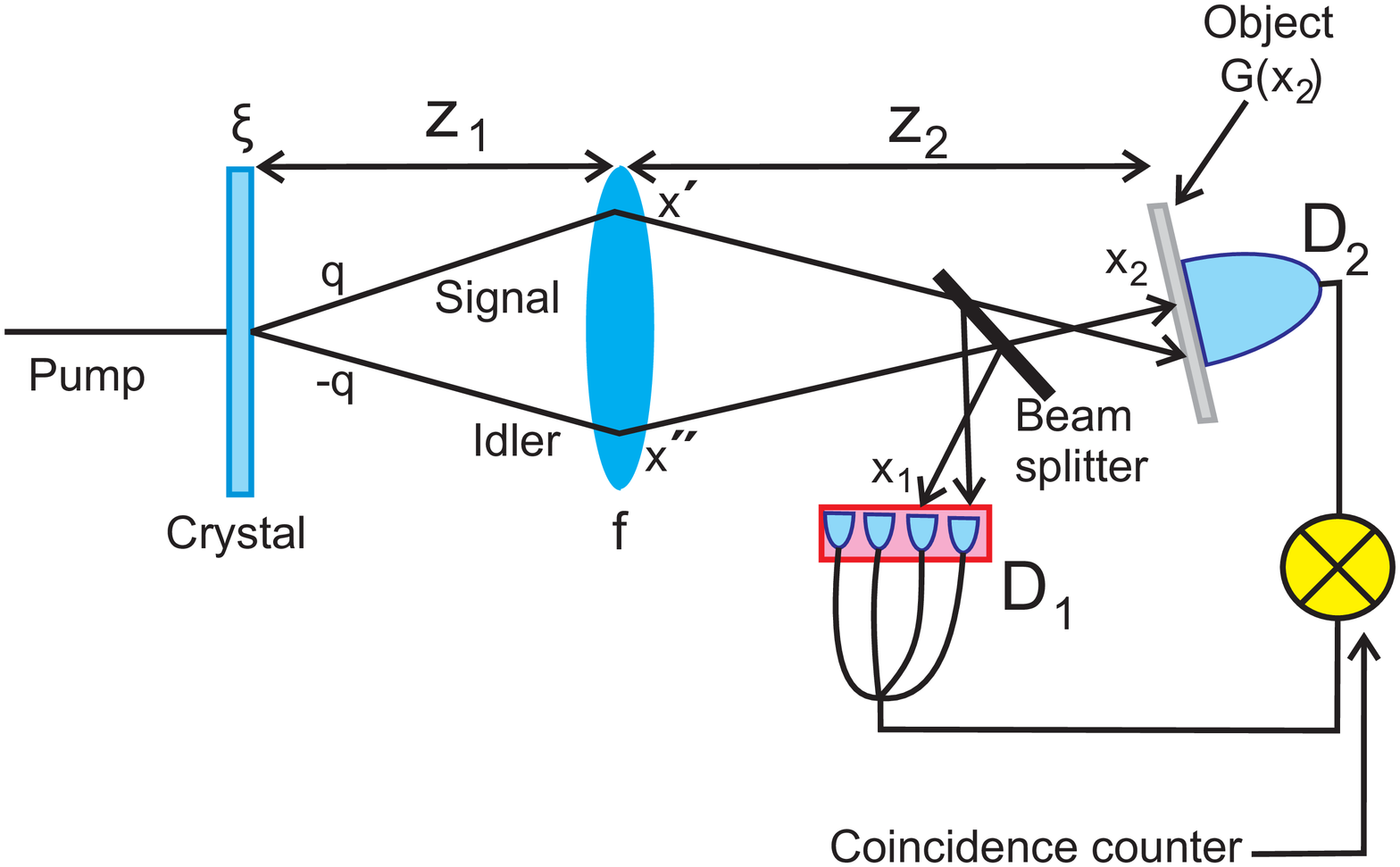}
\caption{\textit{(Color online) Implementation of scheme of fig. 2 using parametric downconversion.}}\label{SPDC}
\end{figure}

Fig. \ref{SPDC} shows a slightly more detailed picture of a particular embodiment of this scheme, using
entangled photon pairs produced via parametric downconversion.
Suppose the field in the pump beam as it enters the crystal is $E_p({\mathbf \xi})$. We
arrange for the distances to satisfy the imaging condition \begin{equation}{1\over {z_1}}+{1\over {z_2}} ={1\over f}.
\end{equation}
The lens will have a pupil function which we assume to consist of the usual quadratic phase plus an additional
phase function describing aberrations: \begin{equation}p({\mathbf x})=e^{-i({\mathbf x}^2/ 2f)}e^{i\phi ({\mathbf x})}.\end{equation} The beam splitter sends half of the photons to a spatially-resolving detector $D_1$, the other
half to a large bucket detector $D_2$, with no spatial resolution.  The object introduces a
modulation $G({\mathbf x_2})$ only in branch 2.
The impulse response function for beam $j$ ($j=1,2$) to travel from transverse position $\xi$ in the crystal plane to transverse position ${\mathbf x_j}$ in the detection plane is then
\begin{eqnarray} h(\xi , {\mathbf x_j})&=&H_j({\mathbf x_j})e^{ik\left[ (\xi^2/z_1)+({\mathbf x_j} /z_2)\right]/2} \\ & & \quad \times \;
\int e^{-ik \left[ ({\mathbf \xi}/ z_1)+({\mathbf x_j}/z_2)\right]\cdot {\mathbf x^\prime}}
e^{i\phi ({\mathbf x^\prime})}d^2x^\prime ,\nonumber\end{eqnarray} where \begin{equation}H_j({\mathbf x_j})=
\begin{cases} 1, & \text{for $j=1$}\\ G( {\mathbf x_2}), & \text{for $j=2$}.\end{cases}\end{equation}

\begin{figure}
\centering
\includegraphics[totalheight=1.8in]{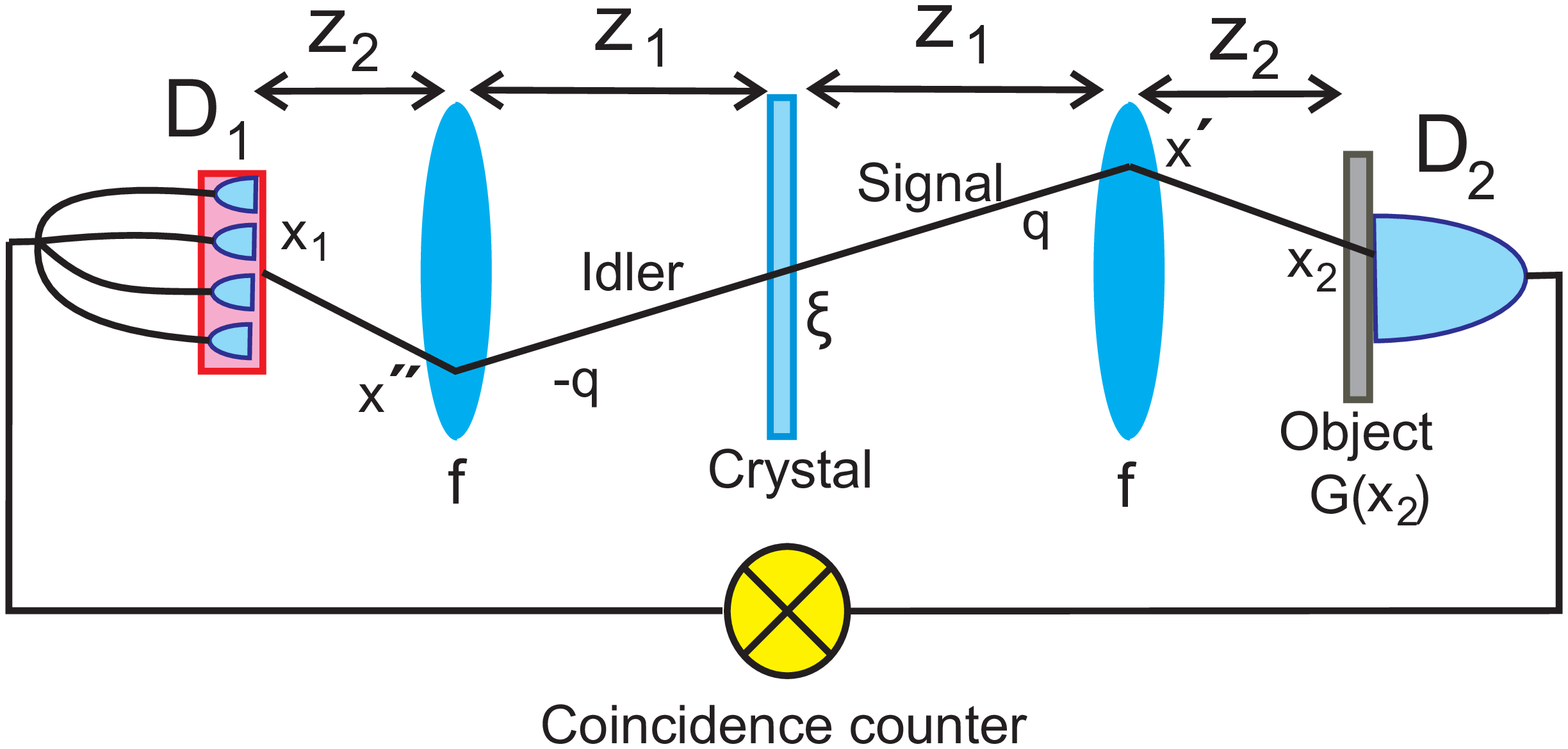}
\caption{\textit{(Color online) Unfolded version of the apparatus in fig. 3, with the signal and idler drawn as a single continuous ray.}}
\label{unfolded}\end{figure}

One further view of the apparatus is useful. Fig. \ref{unfolded} shows an unfolded version of the setup in the Klyshko picture
\cite{klyshko82,klyshko88}. In this picture, the signal and
idler can be viewed as a single continuous ray from one detector to the other. In the approximation of a plane wave pump, this ray is
undeflected at the crystal because of the perfect anti-correlation between the signal and the idler wave vector directions (${\mathbf q} = - {\mathbf q}$), causing the signal and idler to hit the lens at a pair of points arranged symmetrically about the point where the ray crosses the crystal, ${\mathbf x^\prime}-\xi =\xi -{\mathbf x^{\prime \prime}}$.

Assuming a thin crystal and narrow-band filters in the beams, the coincidence amplitude
(or equivalently, the two-photon wavefunction) in the detection plane can be written
(up to overall normalization) in the simple form
\cite{abouraddy}\begin{eqnarray}\psi ({\mathbf x_1}, {\mathbf x_2})&=& \int E_p({\mathbf \xi } )h_1({\mathbf \xi},{\mathbf x_1})
h_2({\mathbf \xi},{\mathbf x_2})d^2\xi \\
&=& \int E_p({\mathbf \xi } ) G({\mathbf x_2 }) e^{(ik\xi^2/ z_1)}e^{(ik /2z_2)
\left( x_1^2+x_2^2\right)}\nonumber \\ & &  \quad \times \;
e^{-ik{\mathbf \xi}\cdot  ({\mathbf x^\prime}+{\mathbf x^{\prime \prime }})/ z_1}
e^{-ik \left( {\mathbf x_1}\cdot {\mathbf x^\prime} +{\mathbf x_2}\cdot {\mathbf x^{\prime\prime}}
\right) } \nonumber \\ & &  \quad \times \;
e^{i\left( \phi ({\mathbf x^\prime })+\phi ({\mathbf x^{\prime \prime} })\right)}d^2\xi d^2x^\prime d^2x^{\prime \prime} . \end{eqnarray}  The coincidence rate is \begin{equation}
R(x_1)= \int d^2x_2 \left| \psi ({\mathbf x_1},{\mathbf x_2})\right|^2 \end{equation}
If we define \begin{equation}F({\mathbf \xi})=E_p({\mathbf \xi})e^{ik\xi^2 /z_1},\end{equation}
then the amplitude may be written in the form \begin{eqnarray}\psi ({\mathbf x_1}, {\mathbf x_2})
&=&G({\mathbf x_2})e^{ik \left( {\mathbf x_2}^2+{\mathbf x_1}^2\right) /2z_2}
\\ & & \times \; \int \tilde F\left({{k \left({\mathbf x^\prime }+{\mathbf x^{\prime\prime}} \right) }\over {z_1}}\right)
e^{-ik \left( {\mathbf x_1}\cdot {\mathbf x^\prime} +{\mathbf x_2}\cdot {\mathbf x^{\prime\prime}}\right)
/z_2 }\nonumber \\
& & \qquad \times \; e^{i\left( \phi ({\mathbf x^\prime })+\phi ({\mathbf x^{\prime \prime} })\right)}
d^2x^\prime d^2x^{\prime\prime}d^2\xi .\nonumber \end{eqnarray}
For simplicity, we take the pump to be an approximate plane wave over the extent of the object, so that we
may set $E_p({\mathbf \xi})$ equal to a constant. We also assume that we are working in the far field, where
the ${\mathbf \xi}^2$ term in the exponential may be neglected in comparison with the other terms. In that case,
we find $\tilde F\left({k\over {z_1}}\left({\mathbf x^\prime }+{\mathbf x^{\prime\prime}} \right)\right)$ is proportional to $\delta^{(2)}\left({\mathbf x^\prime }+{\mathbf x^{\prime\prime}} \right) $, so that
the amplitude reduces to \begin{eqnarray}\psi ({\mathbf x_1}, {\mathbf x_2})
&=&G({\mathbf x_2})e^{ik \left( {\mathbf x_2}^2+{\mathbf x_1}^2\right) /(2z_2) } \\ & & \quad\times \;
\int e^{-ik \left( {\mathbf x_1} -{\mathbf x_2}\right) \cdot {\mathbf x^\prime}/z_2}
e^{2i\phi_{even} ({\mathbf x^\prime })}
d^2x^\prime  \nonumber \\
&=& G({\mathbf x_2})e^{ik \left( {\mathbf x_2}^2+{\mathbf x_1}^2\right)/(2z_2)}\nonumber \\
& & \qquad \times \; \tilde \Phi_{even}
\left( {{k\left( {\mathbf x_1}-{\mathbf x_2}\right)}\over {z_2}}\right) , \end{eqnarray}
where we have defined \begin{equation}\Phi_{even} (x)=e^{2i\phi_{even} ({\mathbf x })},
\label{phieven}\end{equation} and the tilde denotes the Fourier transform.
Given this, the coincidence rate is then
\begin{eqnarray}
R({\mathbf x_1})&=& \int d^2x_2 \left| \psi ({\mathbf x_1},{\mathbf x_2})\right|^2 \\
&=& \int d^2x_2 \left| G({\mathbf x_2}) \tilde \Phi_{even}
\left( {k\over {z_2}}\left( {\mathbf x_1}-{\mathbf x_2}\right)\right)\right|^2 .
\label{mainresult}\end{eqnarray}

Eq. \ref{mainresult} is our main result. Note that it depends only on the even-order aberrations; all odd-order terms have dropped out. In the special case that there is no aberration in the lens, this reduces to
\begin{equation}R({\mathbf x_1})= \left| G\left( {\mathbf x_1}\right) \right|^2 .\end{equation}
It is apparent that we will always obtain unit magnification.

The odd-order aberration cancellation is exact only in the far field and in the case of a plane wave pump. As the
distances involved decrease or as the pump amplitude deviates from a constant, the factor
$\tilde F\left({k\over {z_1}}\left({\mathbf x^\prime }+{\mathbf x^{\prime\prime}} \right)\right)$ will no longer
be a delta function, so that the aberration cancellation will become
only approximate.

The coincidence rate may be compared directly to the image intensity $I(x_1)$ collected by an incoherently-illuminated
single-lens imaging system such as that of fig. 5. Up to overall normalization, the output intensity of the system in fig. 5, taking lens aberrations into account, is \begin{eqnarray}I({\mathbf x_1})&=& \int d^2\xi \left|G({\mathbf \xi })E_p({\mathbf \xi} )\right|^2
\label{ieq1}\\ & & \qquad \times \;\left| \int  d^2 x^\prime
e^{-ik  {\mathbf x^\prime }\cdot \left( (\xi /z_1) +(x_1/z_2)\right)}
e^{i\phi ({\mathbf x^\prime })}\right|^2\nonumber \\
&=& \int d^2\xi \left|G({\mathbf \xi })E_p({\mathbf \xi} )\tilde \Phi \left(k\left( {{\mathbf \xi}\over {z_1}}
+{{\mathbf x_1}\over {z_2}}\right)\right) \right|^2,\label{ieq2}
\end{eqnarray} where \begin{equation}\Phi ({\mathbf x })=e^{i\phi ({\mathbf x })}.\label{phi}\end{equation}

\begin{figure}
\centering
\includegraphics[totalheight=1.7in]{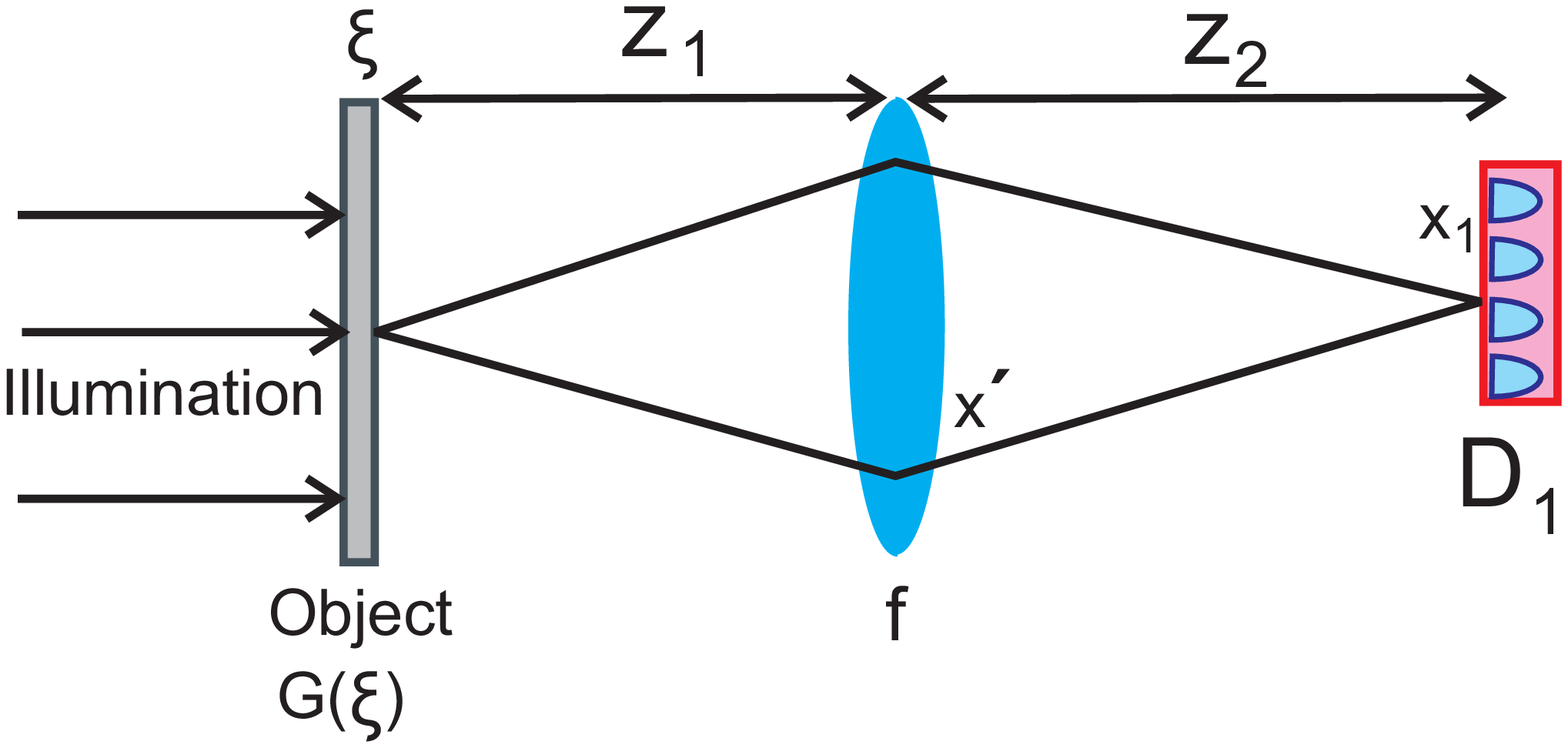}
\caption{\textit{(Color online) Simple incoherent imaging system. The distances $z_1$ and $z_2$ obey the imaging condition
${1\over {z_1}}+{1\over {z_2}}={1\over f}$.}}
\end{figure}

Setting $E_p({\mathbf \xi} )=$constant (plane wave illumination),
we see that in the aberration-free case ($\phi({\mathbf x_1}) =0$) both $\tilde \Phi$ and $\tilde \Phi_{even}$
become delta functions, so that eqs. \ref{mainresult} and
\ref{ieq2} both lead to images of the form $|G(M{\mathbf x_1})|^2$, where $M=+1$ for eq.
\ref{mainresult} and $M=-\left( {{z_2}\over {z_1}}\right)$ for eq. \ref{ieq2}.
%Henceforth, we take $z_1=z_2$ for the sake of comparison.

The advantage of the proposed setup is clear in the case where the aberrations are all of odd order:
in this case, the factor of $\tilde \Phi $ in
eq. \ref{ieq2} distorts the image, whereas the image formed via eq. \ref{mainresult} is unaffected.
However, the price to be paid for this is seen by considering the case where the aberrations are entirely even
order: the factor of $2$ in the exponent of eq. \ref{phieven} doubles the effect of the even-order aberrations
(compare equation (\ref{phieven}) to eq. (\ref{phi})).
The reason for this is clear: both the signal and the idler contribute to the image and both gain extra phases from the aberration. For the odd-order terms, the two phases cancel, while in the even-order terms the phases add constructively. So, while the setup proposed here eliminates odd-order aberrations, it does so at the expense of worsening even-order aberrations. This method will therefore be of maximal benefit when the aberration term of greatest
importance is of odd order and is much larger than any of the even order terms.

A further advantage is the improvement in sensitivity over standard single-detector imaging due to reduced effects of noise. For example, we can look at the effects of the detector dark current on the coincidence rate. We assume (i) that all fluctuations have zero mean, and (ii) the dark current fluctuations in one detector are uncorrelated with the signal and dark currents in the other detector. Then we can write the total detected current as the sum of the signal and dark currents ($I_s$ and $I_d$), and (because of assumption (i)) further split each signal and dark current into mean and fluctuation parts: \begin{equation} I_j=I_{sj}+I_{dj}=
\langle I_{sj}\rangle +\langle I_{dj}\rangle +\delta I_{sj}+\delta I_{dj},\label{currents}\end{equation} where $j=1,2$ labels the detector.
If the average background counting rate is subtracted off, the remaining coincidence rate is proportional to the correlation function between the measured currents in the two detectors, \begin{equation}G^{(2)}(I_1,I_2)=\langle I_1I_2\rangle -\langle I_1\rangle \langle I_2\rangle .\label{G2}\end{equation} Substituting eq. \ref{currents} into eq. \ref{G2} and making use of the two assumptions above, a few lines of algebra quickly shows that the effect of the dark current completely cancels out of the correlation function.

The effect of quantum noise is more complicated, and has been studied recently in a number of papers \cite{cheng,gatti2,saleh, erkmen}. Since the effect of quantum noise is strongly dependent on the parameters of the experiment, it is hard to draw general conclusions about its relative effect on quantum versus classical imaging. In some cases quantum imaging offers benefits on this front as well, but the situation has to be examined on a case by case basis. See the references cited above for more detail.

\section{Aberration-Cancellation with a Classical Source}

The odd-order aberration cancellation effect in the previous section occurs
because the beams strike the lens in a spatially anticorrelated manner, leading to the
structure $\phi ({\mathbf x^\prime})+\phi ({\mathbf x^{\prime \prime}})
=\phi ({\mathbf x^\prime})+\phi (-{\mathbf x^\prime })$ in the exponents.
The entanglement of the beams plays no role here.
The same effect may be produced by any method that requires light to strike diametrically
opposite points on the lens simultaneously.

\begin{figure}
\centering
\includegraphics[totalheight=1.9in]{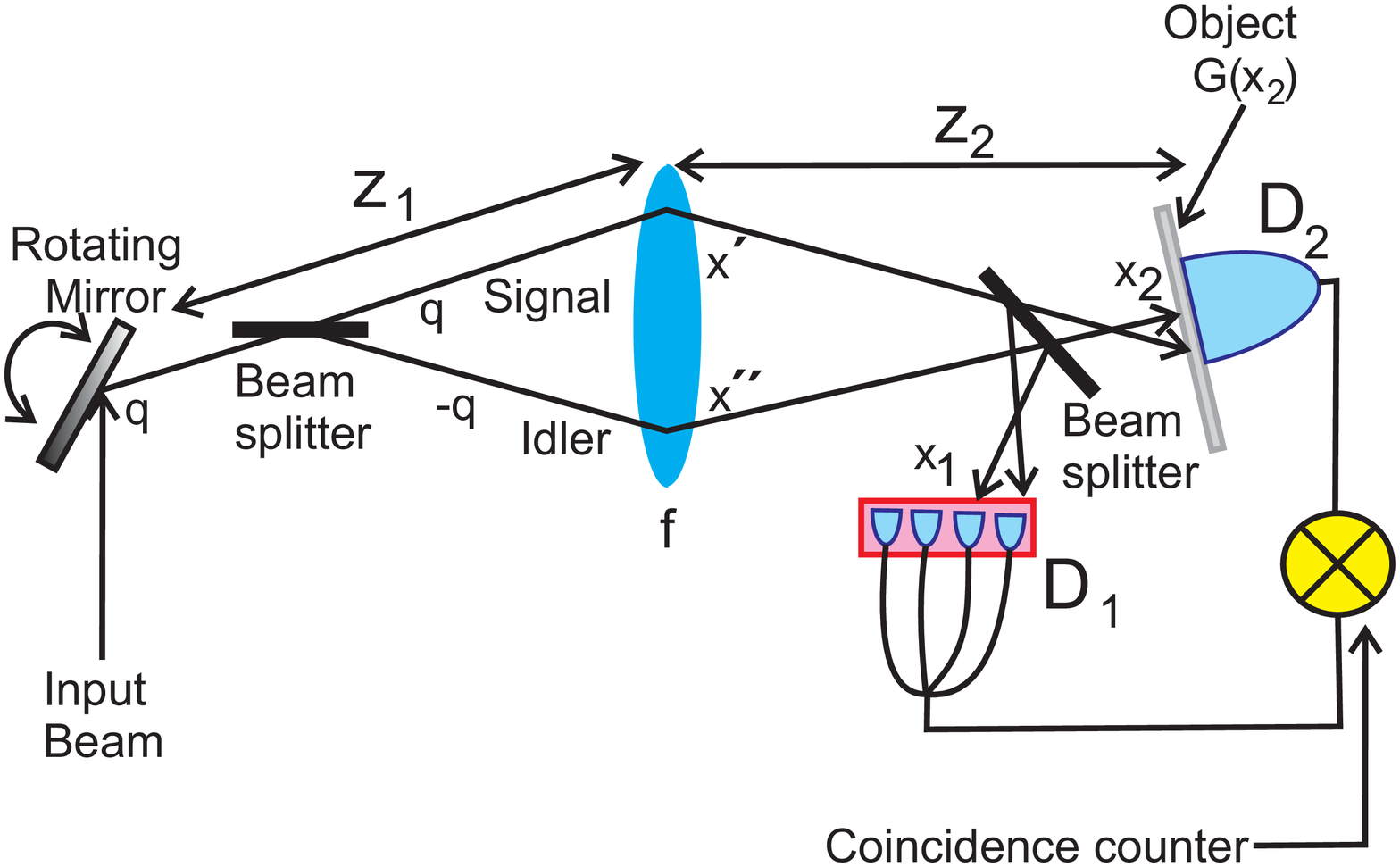}
\caption{\textit{(Color online) Implementation of scheme of fig. 2 using narrow classically correlated beams for illumination.}}
\label{classical}\end{figure}

To illustrate this, we display in fig. \ref{classical} an apparatus with classical illumination that achieves the
same effect. A narrow beam illuminates a rotating mirror, which
reflects the beam onto a beam splitter. (The rotating mirror could be replaced by any form of
beam-steering modulator.) The reflected and transmitted beams
leave the beam splitter with opposite transverse momenta $\pm {\mathbf q}$, striking the lens
on opposite sides of the axis. Over time, as the mirror rotates, the entire area of the lens is filled.
As before, the object is placed in front of the bucket detector $D_2$. The
impulse response function for the full system is the product of those of the two individual beams, once again
leading to the same $\phi ({\mathbf x^\prime})+\phi (-{\mathbf x^\prime})$ structure, and to odd-order
aberration cancellation.

\section{Conclusions}

In this paper, we have shown that correlated photon methods may be used to manipulate the effects of aberration in a simple optical imaging system. Specifically, the scheme described here will eliminate effects of odd-order aberrations induced by the optical system at the expense of amplifying the effects of even-order aberrations. The results here are in a sense complementary to those of \cite{bonato1,bonato2,simon1}, which involved cancellation of {\it even}-order aberrations induced by the object itself (not by the optical system) in interferometry by means of entangled photons. It remains for future investigations to see what additional types of manipulation of
phase effects may be possible using variations on the methods considered here.

\begin{acknowledgments}

This work was supported by a U. S. Army Research
Office (ARO) Multidisciplinary University Research Initiative
(MURI) Grant; by the Defense Advanced Research Projects Agency (DARPA),
and by the Bernard M. Gordon Center for Subsurface Sensing and Imaging
Systems (CenSSIS),an NSF Engineering Research Center.

\end{acknowledgments}

\end{document}